\documentclass[12pt]{article}
\usepackage{graphicx}
\begin{document} 
\baselineskip 18pt

\bigskip
\centerline{\bf \large Simulated self-organisation of death by inherited mutations}

\bigskip
\noindent
J.S. S\'a Martins$^{1,2}$, D. Stauffer$^{1,3}$, P.M.C. de Oliveira$^{1,2}$, 
and S. Moss de Oliveira$^{1,2}$

\bigskip
\noindent
$^1$ Laboratoire PMMH, \'Ecole Sup\'erieure de Physique et de Chimie
Industrielles, 10 rue Vauquelin, F-75231 Paris, France

\medskip                
\noindent
$^2$ Visiting from Instituto de F\'{\i}sica, Universidade
Federal Fluminense; Av. Litor\^{a}nea s/n, Boa Viagem,
Niter\'{o}i 24210-340, RJ, Brazil

\medskip
\noindent
$^3$ Visiting from Institute for Theoretical Physics, Cologne University,
D-50923 K\"oln, Euroland

\bigskip

\bigskip
{\small 
An agent-based computer simulation of death by inheritable mutations in a
changing environment shows a maximal population, or avoids extinction, at some 
intermediate mutation rate of the individuals. Thus death seems needed to allow 
for evolution of the fittest, as required by a changing environment.
}

\bigskip

\section{Introduction}
More than a century ago Weissmann argued that ageing and death are needed to
make place for our children; and children are in turn needed to allow for 
Darwinian evolution through survival of the fittest. Kirkwood \cite{kirk}
summarized this and many other theories of ageing, and specific computer models
of ageing and death are reviewed e.g. in \cite{book1,book2}, for example the 
Penna model (as also reviewed in \cite{penna}). (A mathematical argument 
against immortality was given in this sense in \cite{pmco}.)

Now we want to understand the need for death through Monte Carlo simulations
of individuals. We distinguish between newborns and adults, and take 
into account environmental changes. They may come from climate change, like 
ice ages and warmer periods during the existence of {\it homo sapiens}. Or they
may be caused by migrations of people from one environment to another. A single
such environmental change was already used to justify sexual over asexual 
reproduction \cite{sex}. Thus we vary the mutation rate of individuals to find 
its optimal value. Here "optimum" either means a maximum of the population
in a fixed environmental carrying capacity, or survival instead of extinction, 
depending on which of our two models A and B we use.

In our two models A and B, using sexual reproduction, the genome is represented 
by two strings of $L$ bits each. They represent the $L$ most serious 
genetic diseases. Each mutation damaging the phenotype (i.e the health of 
the individual) reduces the survival probability per iteration by a factor 
$x$. As genetic load we count those bit positions which differ from an ideal 
bit-string. The latter is initially zeroed, but changes at each iteration with a probability $p$ at one
randomly selected bit position, and thus represents the changing environment. 
For reproduction the two bit-strings of the father are crossed-over at one randomly 
selected bit position, the same happens for the mother, and then one of the two 
resulting bit-strings from the father (the gamete) is combined with one of the two 
from the mother to give the child genome. 
Mutations are thus inherited from the parents, and $m$ new mutations are 
introduced at birth to each gamete (if $m \ge 1$; for $m < 1$, one new mutation is added with 
probability $m$).
$N$ is called the genetic load; more precisely it is the number of 
loci (bit positions) where the genome is not adapted to the current environment.
All changes in the individuals and the environmental bit-strings are reversible.

Model A, discussed first, uses a varying population and finds as an optimal
$m$ that mutation rate for which the equilibrium population reaches a maximum. 
Model B, discussed thereafter, follows a tradition of theoretical biology and 
keeps the population constant except if during one iteration all adults die out;
then we check which mutation rate avoids extinction of the whole population. 
Further details of the two models will be discussed in the corresponding
sections.

\begin{figure}[!hbt]
\begin{center}
\includegraphics[angle=-90,scale=0.4]{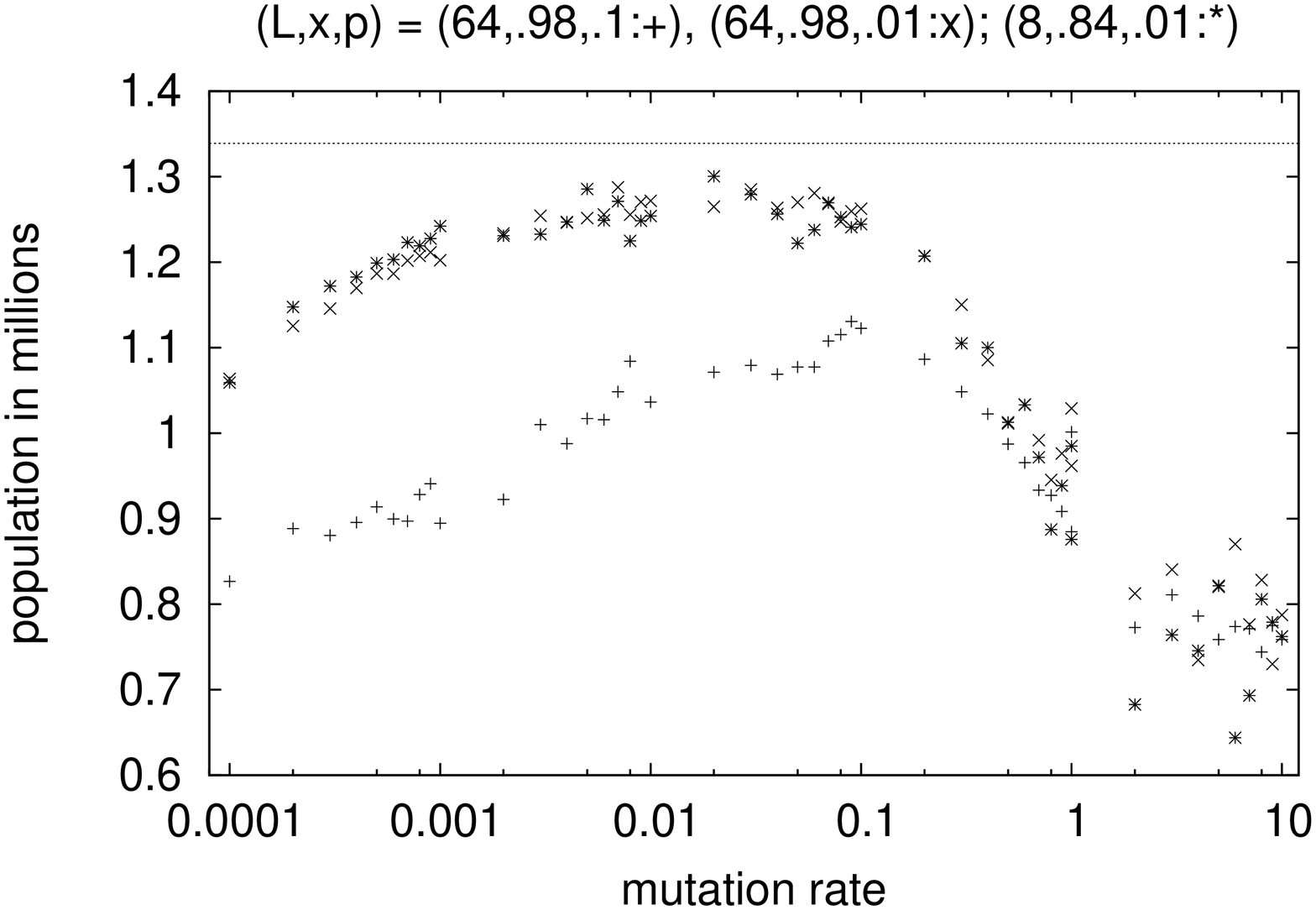}
\includegraphics[angle=-90,scale=0.4]{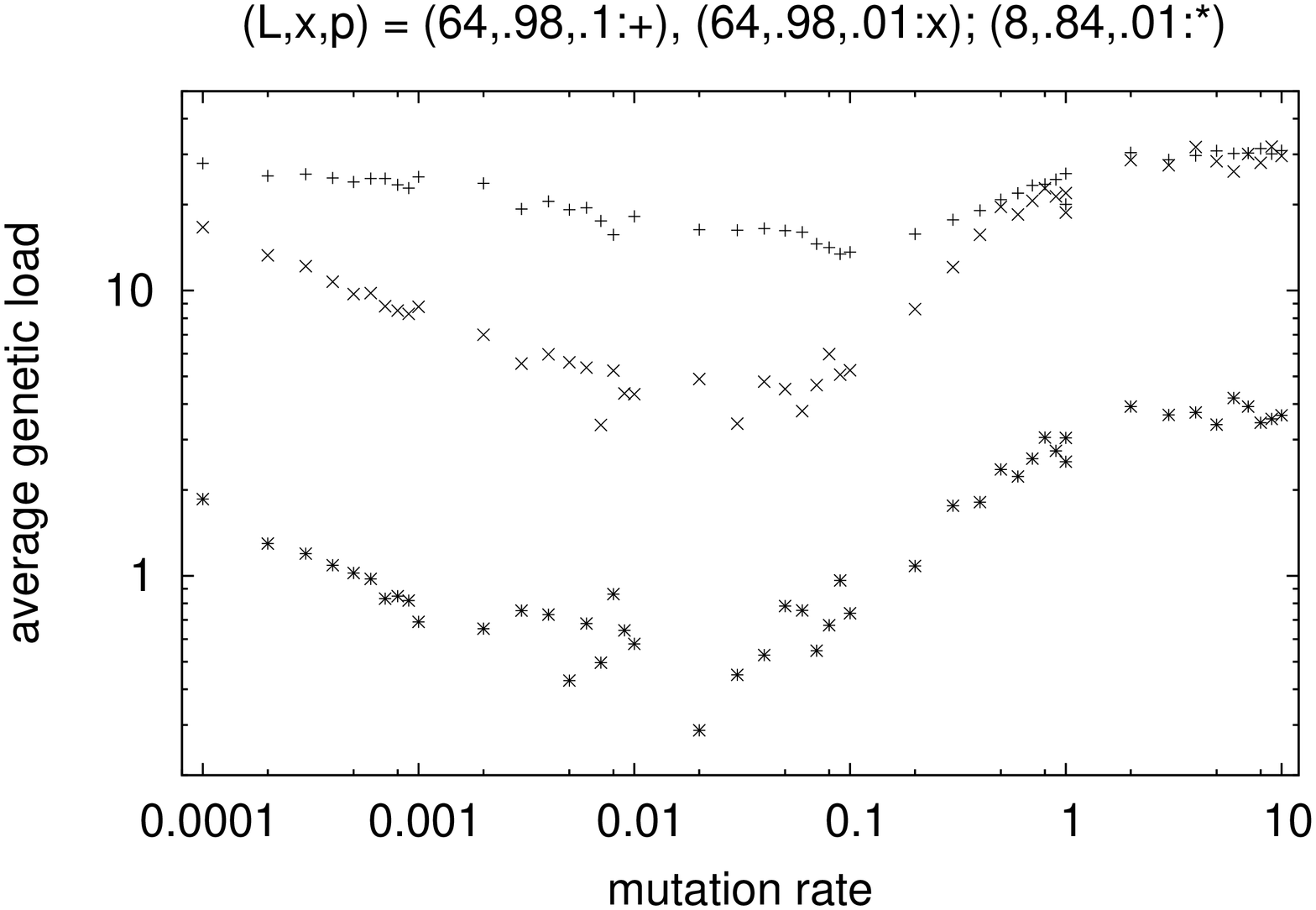}
\end{center}
\caption{Model A1. Search for the optimal mutation rate, where the population
(top) gets maximal and the number $<N>$ of unadapted loci (genetic load, bottom) 
gets minimal, at $x=0.98$. For $x = 0.99,\, L=64, \, p=0.01$ the results are
similar, for $x=0.96$ the populations die out for some of these parameters.
}
\end{figure}

\begin{figure}[!hbt]
\begin{center}
\includegraphics[angle=-90,scale=0.5]{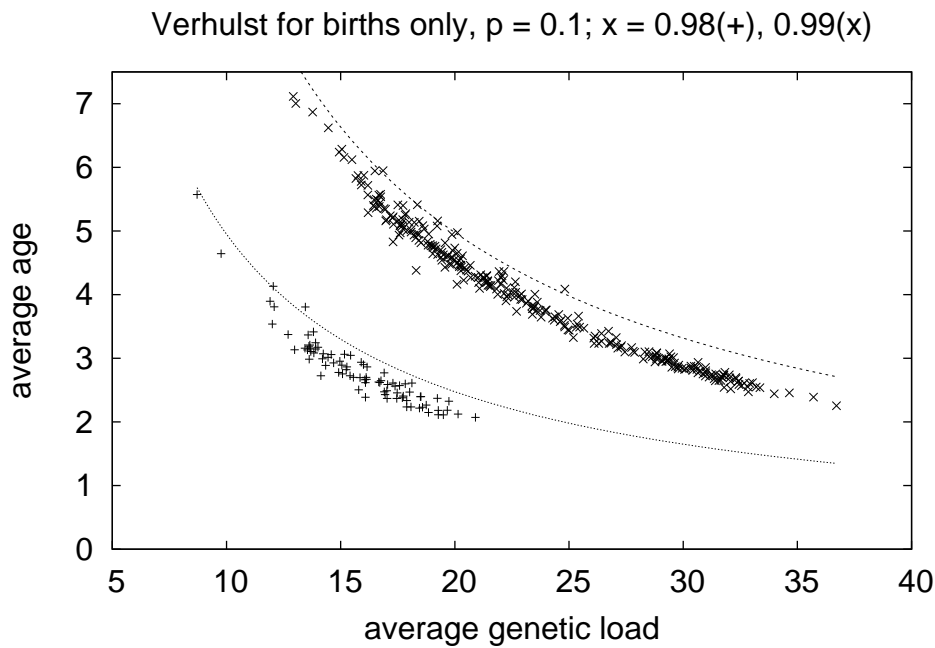}
\end{center}
\caption{Model A1. Average age of survivors versus number of unadapted loci,
when the Verhulst death probability applies to the births only; 64 bits, 
various $m$ and various observation times. The curves show $1/(|\ln x| N)$.
}
\end{figure}

\begin{figure}[!hbt]
\begin{center}
\includegraphics[angle=-90,scale=0.4]{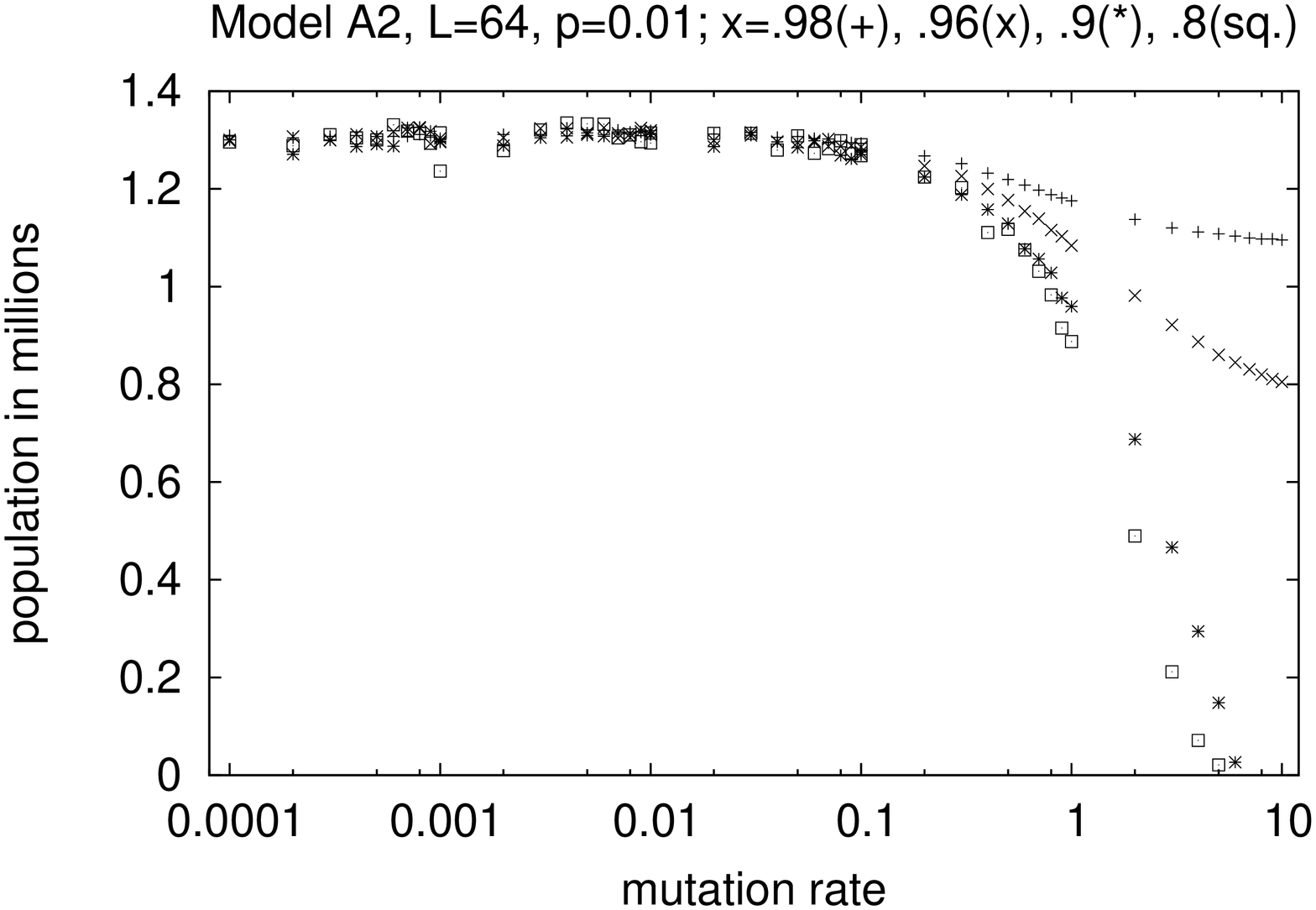}
\includegraphics[angle=-90,scale=0.4]{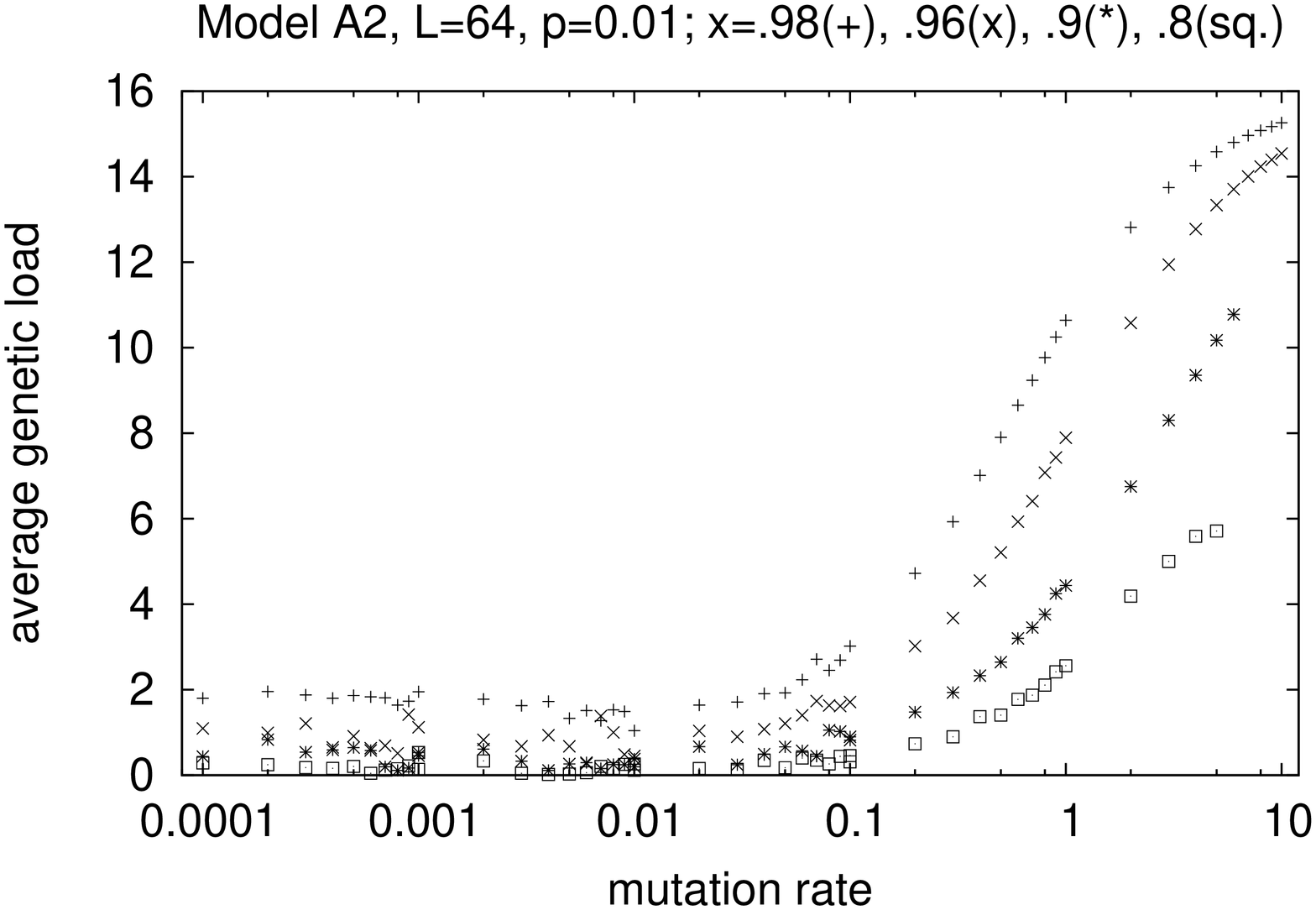}
\end{center}
\caption{Model A2. As Fig.1 but with modified recessiveness and smaller $x$.
}
\end{figure}

\section{Model A with changing population}

In model A, each of the individuals survives the next time step (iteration
involving all survivors) with probability $x^N(1-P/K)$ where $P$ is the current
total population and $K$ is a fixed input parameter, sometimes called the
carrying capacity representing limitations (due to lack of food and space)
for the growth of the population. Here $N$ bits are not adapted to the current 
environment. (The Verhulst factor $1 - P/K$ applies to all individuals, 
differently from \cite{cebrat}.)

Our mutations are recessive, which is defined differently
in two different versions A1 and A2 of model A. For A1 we take the logical
$and$ of the two bit-strings of the individual,  and then count as $N$ the 
number of bit positions where this logical $and$ differs from the current ideal
bit-string. This version is close to \cite{stauffercebrat}, and means that heterozygous 
loci do not count for the genetic load $N$ \underline{before} the first environmental 
change, but count \underline{after} it. For A2 we count for 
$N$ only those positions where both individual bit-strings agree with each 
other (homozygous loci) and disagree with the ideal bitstring.

Our $K$ is mostly 2 million, the initial population is $K/5$, and the 
resulting equilibrium population is mostly of the order of one million if 
it does not die out. The two individual bit-strings are mutated independently,
each with mutation rate $m$. Each surviving adult at each iteration gives birth to $B$ babies,
which become adult at the next iteration; we used $B=4$. Mostly 10,000 
iterations were made (100,000 for most cases with $m < 0.001$), and averages 
were taken from the second half of this time interval.  

{\it Case A1:}
Figure 1 shows our main result: The population $P$ has a maximum as a function 
of $m$ at some intermediate $m$ value. Thus neither very small $m$ ("eugenics")
nor very large $m$ ("instability") are optimal; an intermediate mutation
rate leads to the largest $P$ or the lowest $<N>$, but also in reality to a 
finite lifespan. 

Instead of applying the Verhulst deaths to all ages, Fig.2 shows the 
correlation between {\it genetic} deaths only and genetic load, by applying the
Verhulst deaths only to the births \cite{cebrat}. Data are taken from averages 
calculated at different time steps of a single run for each value of $x$.
 
{\it Case A2:}
The modified recessiveness defined above for model A2 reduces $N$ and makes 
survival possible for a changing environment even for unrealistically small $x$.
For $x \ge 0.8$ we see in Fig.3 top  a plateau for small mutation rates $m$, 
followed by a decay for larger $m$. Thus in this less realistic case A2 there 
is no longer the clear population maximum as it was seen in model A1. A similar 
result was obtained for model A1 in a stable environment (not shown).

\section{Model B with constant population}

In this version there are no random deaths by the Verhulst factor. Instead, individuals 
die exclusively due to genetic reasons. At each time-step, each individual with a genetic 
load $N$ survives with probability $x^{N+1}$. So, an individual with zero load can still die 
with probability $1 - x$. This selection mechanism may lead to the extinction of the whole 
population, for some values of the model's parameters. If there is no extinction at a given 
time step, the survivors breed, generating new individuals until the initial population size 
is restored for the next time step. There is no distinction between males and females, and 
the population may be regarded as one of hermaphrodites.

To generate the offspring's genome, the genetic strings of each parent are crossed-over and one 
gamete of each is randomly chosen. A number $M$ of mutations extracted from a uniform 
distribution in the interval $[0,2m)$ is then introduced in this genome, each one at a random 
location of a randomly chosen gamete. Thus, $m = 2$ in this model (B) corresponds to $m = 1$ 
in model A. If $M$ is not an integer, then $int(M)$ mutations are 
added, where $int(x)$ is the largest integer contained in $x$, and an extra mutation is added 
with probability $M - int(M)$. As a result of this strategy, $m$ new mutations are added to 
each offspring genome on average.

The model treats heterozygous loci in the same way as model A2, that is, they never contribute 
to the genetic load. A slightly different version of this model, in which $x$ was recalculated 
at each time step to keep constant the fraction of deaths, was presented in Ref. \cite{pm}. 

The results for this model shown in Fig. 4 should be compared to our Fig. 1. We keep 
the mutation rate of the environment $p = 0.01$ and the selection strength $x = 0.98$ fixed 
and compute the average genetic load $<N>$, the fraction 
of the population that dies per time step, and the fraction of perfect, or ideal, genomes in 
the population as the mutation rate $m$ is varied. We find that there is an intermediate range 
of values of the mutation rate for which both the genetic load and the death rate go through minima, 
while the fraction of perfects reaches a maximum. This result matches what was found, in similar 
situations, in our model A1.

Our main result refers to the need for a strong selection mechanism as a means to enforce a 
small genetic load: death of the least adapted individuals makes way to fitter ones. In Fig. 5 
we show the time evolution of the average genetic load of the population for four 
different sets of parameters. In all four, we simulate a population of $1000$ individuals, each 
represented by two bit-strings of size $3200$ bits each, with a mutation rate at birth of $m = 1.0$. 
In case (a), $x = 0.98$ (weak selection) and $p = 0$, the environment does not change. The average 
genetic load starts at $0$ (ideal individuals) and grows to a small value of order $1$. The 
distributions of genetic loads are shown in Fig. 6, averaged after the initial $5000$ time 
steps. For case (a), it has a peak at $0$, meaning that a majority of the population carries no 
genetic load, with a small width. When the environment changes with 
probability $p=0.01$ at each time step (case (b)), the average genetic load increases to a value of 
order $10$ and its distribution peaks at a small non-zero value of the same order. Further increase 
in the rate of environment change to $p=0.02$ leads the population to extinction (case (c)). The 
average genetic load increases rapidly and its distribution widens: it is shown in Fig. 6. 
The genetic load accumulates thanks to the joint effects of the mutation 
rate at birth and a fast environment change and, even with a weak selection, leads eventually to 
extinction. The need for a strong selection is now shown: for the same parameters ($p$ and $m$) but 
smaller $x = 0.95$ (case (d)), the population resists and the distribution of genetic load is very 
similar to the one in case (b).

The same qualitative results were obtained for haploid asexual 
populations, but extinction is avoided only for larger populations and stronger selection 
pressure, similar to \cite{asex}.

Extinction can then be correlated to features of the distribution of genetic load. It is avoided 
as long as the average genetic load is not much larger than the width of the distribution. This is more clearly 
shown in Fig. 7, where we plot the results of simulations of populations represented by two 
bit-strings of size $2048$ bits each, with $x = 0.9$ (strong selection), $m = 1$, and $p$ is varied in the 
interval $[0, 0.36]$. Both the average genetic load and the width of the distribution increase 
monotonically with $p$, while the fraction of individuals with zero load decreases. Beyond $p = 0.35$ 
the latter vanishes and extinction is the outcome of the simulation. In the same plot 
we also show the fraction of individuals that die (for genetic reasons only in this model) at each time 
step. As $p$ is increased, survival of the population becomes more difficult and causes this fraction 
to be ever increasing.

\begin{figure}[!hbt]
\begin{center}
\includegraphics[angle=-90,scale=0.4]{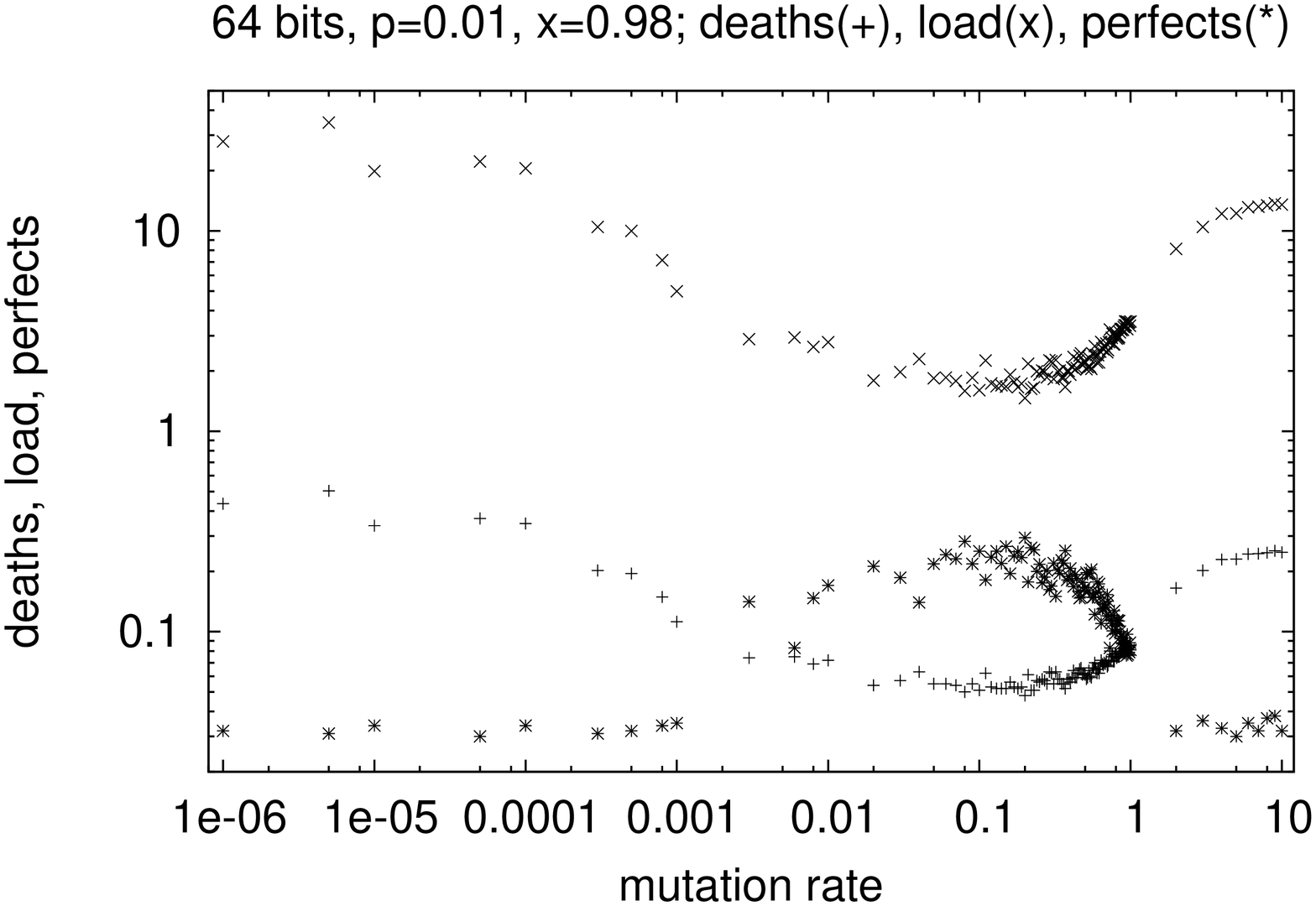}
\includegraphics[angle=-90,scale=0.4]{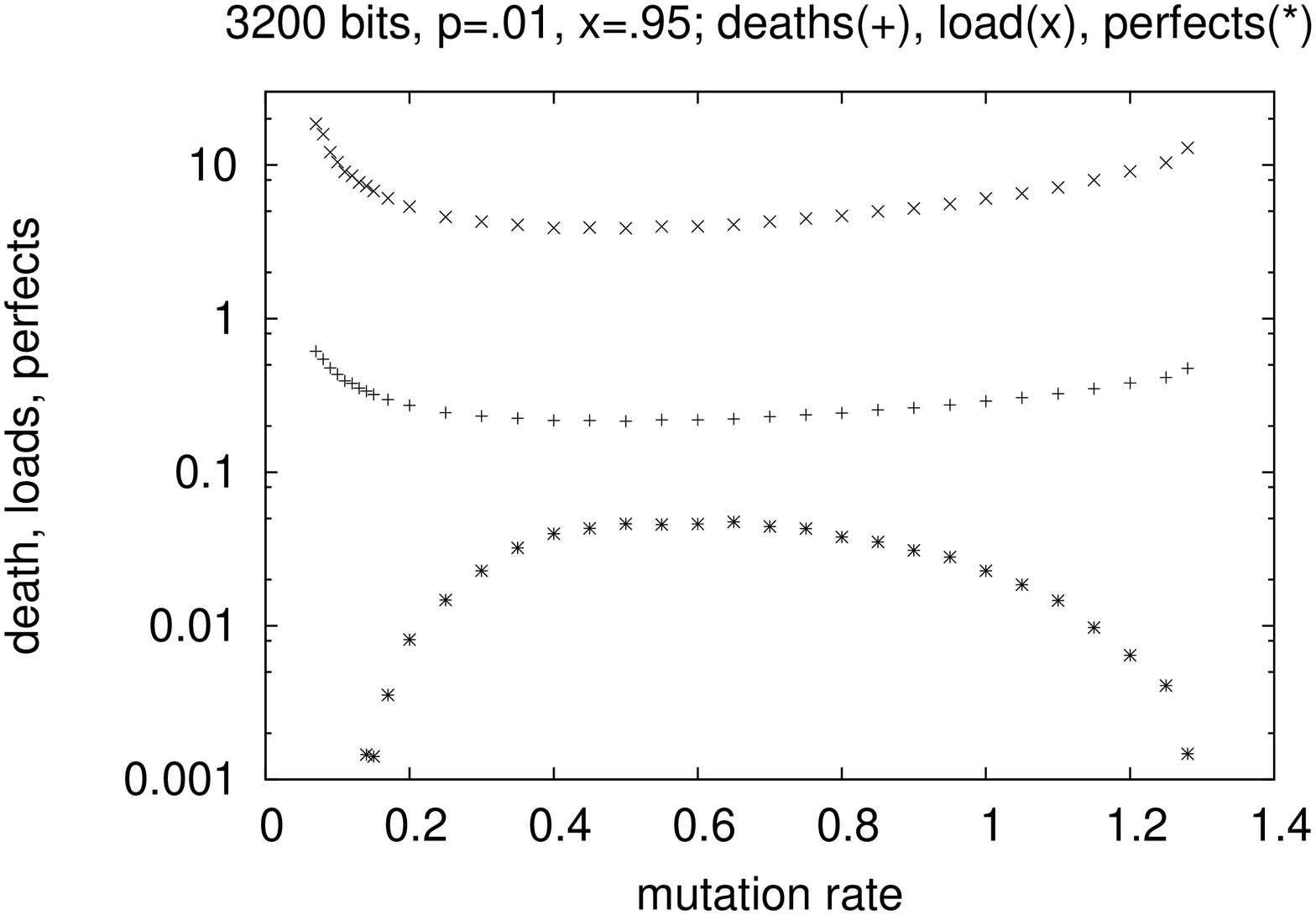}
\end{center}
\caption{Average genetic load $<N>$, deaths and number of perfect individuals as a fraction of the population. 
Top figure for bit-strings of size $64$, with simulations running for $10^6$ time steps, and bottom figure 
for bit-strings of size $3200$.}
\label{B64}
\end{figure}

\begin{figure}
\begin{center}
\includegraphics[angle=-90,scale=0.4]{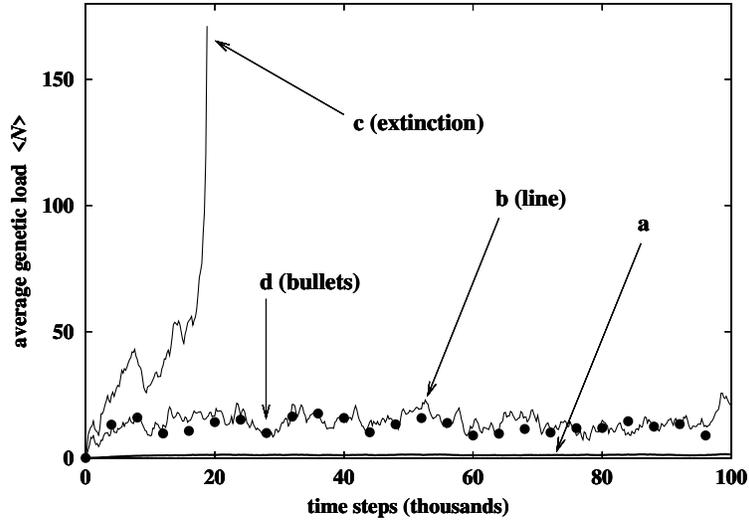}
\end{center}
\caption{Time evolution of average genetic load for the four cases ((a)-(d)) described in the text. Case (c) 
leads to extinction, while case (d) shows survival when selection is increased.}
\label{B1}
\end{figure}

\begin{figure}
\begin{center}
\includegraphics[angle=-90,scale=0.4]{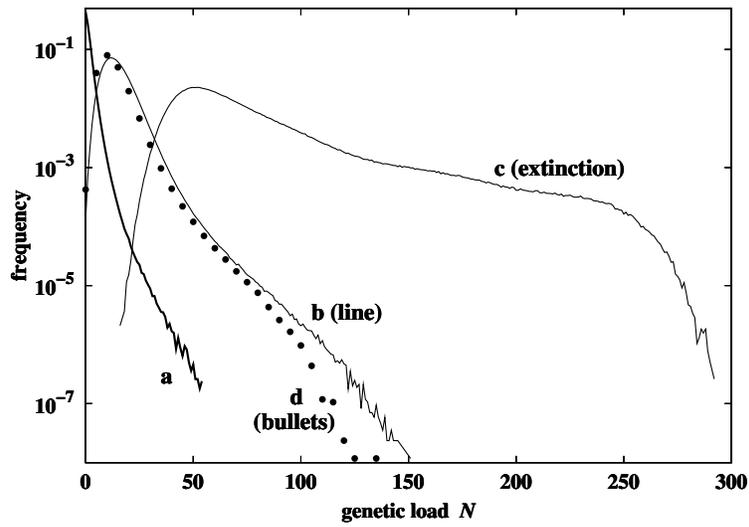}
\end{center}
\caption{Distribution of genetic load for the four cases ((a)-(d)) described in the text.}
\label{B2}
\end{figure}

\begin{figure}
\begin{center}
\includegraphics[angle=-90,scale=0.4]{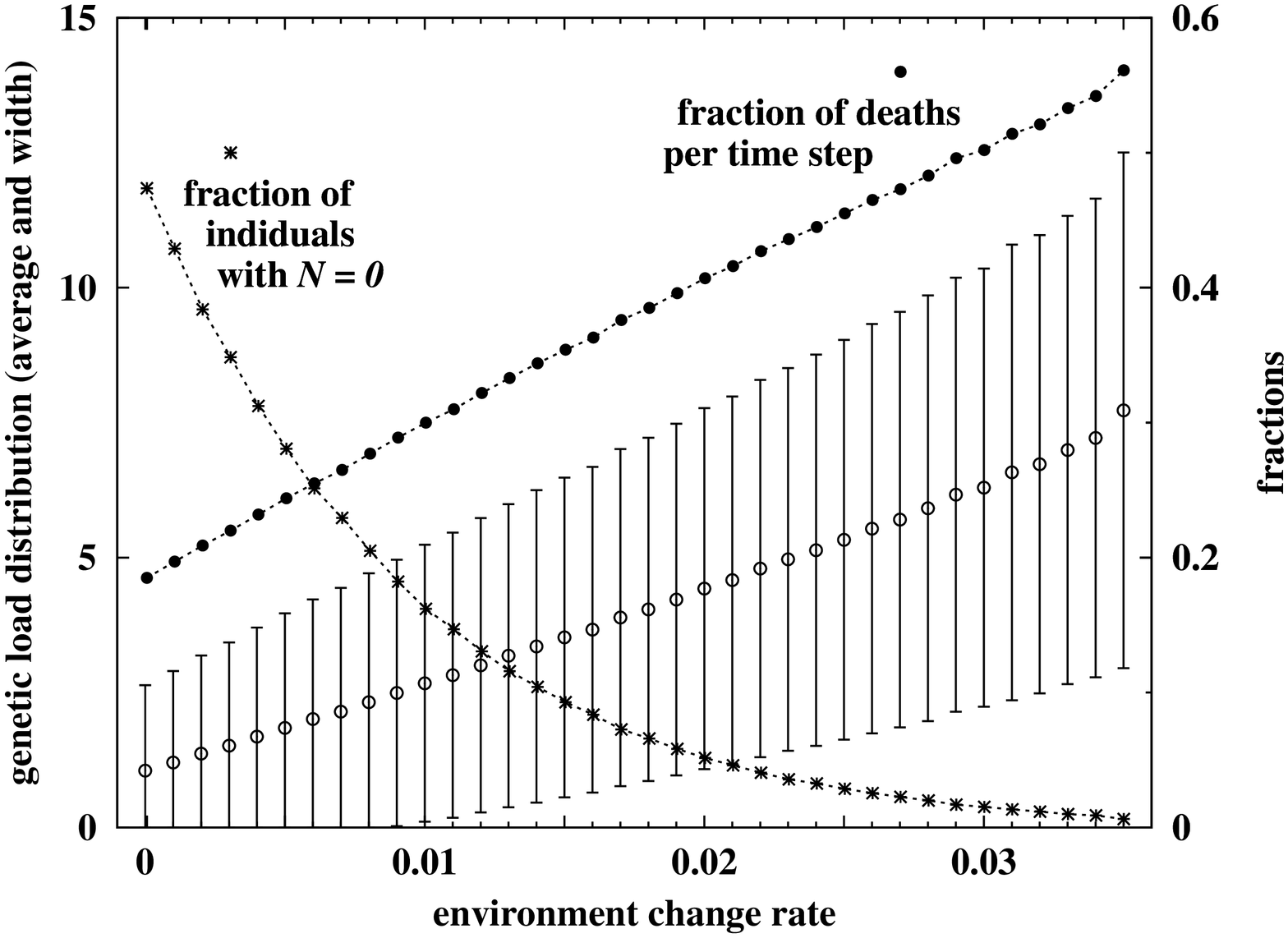}
\end{center}
\caption{Average genetic load - open circles - and width of the distribution of load - error bars - 
(y scale on the left), fraction of the population with no genetic load and fraction of deaths per 
time step (y scale on the right). Values correspond to averages taken after $5000$ initial time 
steps, up to $10^6$. Similar results are obtained for $L \le 32768$.}
\label{B3}
\end{figure}

\section{Conclusion}

In all our models, the genetic heritage of a diploid individual is represented 
by a pair of bit-strings, which undergo mutations at birth, while the ideal phenotype 
is mapped into a single bit-string. 
Environmental change is translated into a mutation of this ideal phenotype. The genetic 
load of an individual is determined by a comparison between its genetic strains and 
the ideal phenotype. This genetic load determines the death probability of each individual. 

Our results come from simulations with a fixed rate of environment change and a fixed 
value for the parameter that measures selection strength $x$. 
We show that population fitness, determined by its size, reaches a broad maximum, while the 
average genetic load reaches a minimum, for some intermediate range of the mutation rate at 
birth (model A1). 
So, nature has self-organised its cellular error correction machinery to ensure a mutation rate 
within some range.

On the other hand, when the rate of environment change increases, our results are 
consistent with the interpretation that selection has to get stronger to avoid population 
extinction (model B).

A more realistic approach would be perhaps to assign a different selective value for each 
different bit position, since different inherited diseases differ in their danger to survival. 
However, that modification would introduce so many free parameters that the model would lose 
its value.

\end{document}